\documentclass{article}
\usepackage{LaThuileFPSpro,psfig}
\begin{document}
\title{ 
  NEUTRINO ASTRONOMY AT THE SOUTH POLE \\
  STATUS OF THE AMANDA EXPERIMENT
  }
\author{
  Paolo Desiati        				\\
  {\em desiati@amanda.wisc.edu} 		\\
  {\em Physics Department, University of Wisconsin, Madison, U.S.A.}\\
  for the AMANDA Collaboration        		\\
  }
\maketitle
%
% add sloppypar command to prevent hyphenation
%
{\scriptsize
\begin{sloppypar}

\noindent
J.~Ahrens$^{11}$, 
X.~Bai$^{1}$, 
S.W.~Barwick$^{10}$, 
T.~Becka$^{11}$, 
J.K.~Becker$^{2}$,
K.-H.~Becker$^{2}$, 
E.~Bernardini$^{4}$, 
D.~Bertrand$^{3}$, 
A.~Biron$^{4}$, 
D.J.~Boersma$^{4}$
% ...
\end{sloppypar}

\vspace*{0.5cm} 

{\it
\noindent
   (1) Bartol Research Institute, University of Delaware, Newark, DE 19716, USA
   \newline
   (2) Fachbereich 8 Physik, BUGH Wuppertal, D-42097 Wuppertal, Germany
   \newline
   (3) Universit\'e Libre de Bruxelles, Science Faculty, Brussels, Belgium
   \newline
   (4) DESY-Zeuthen, D-15738 Zeuthen, Germany
   \newline
   (5) Blackett Laboratory, Imperial College, London SW7 2BW, UK
   \newline
   (6) Dept. of Technology, Kalmar University, S-39182 Kalmar, Sweden
   \newline
   (7) Lawrence Berkeley National Laboratory, Berkeley, CA 94720, USA
   \newline
   (8) Dept. of Physics, Pennsylvania State Univ., University Park, PA 16802, USA
   \newline
   (9) Dept. of Physics, University of California, Berkeley, CA 94720, USA
   \newline
   (10) \hbox{Dept. of Physics and Astronomy, Univ. of California, Irvine, CA 92697, USA}
   \newline
   (11) Institute of Physics, University of Mainz,  D-55099 Mainz, Germany
   \newline
   (12) Dept. of Physics, University of Maryland, College Park, MD 20742, USA
   \newline
   (13) University of Mons-Hainaut, 7000 Mons, Belgium
   \newline
   (14) Dept. de F\'{\i}sica, Universidad Sim\'on Bol\'{\i}var, Caracas, 1080, Venezuela
   \newline
   (15) Dept. of Physics, University of Wisconsin, Madison, WI 53706, USA
   \newline
   (16) Physics Dept., University of Wisconsin, River Falls, WI 54022, USA
   \newline
   (17) Div. of High Energy Physics, Uppsala University, S-75121 Uppsala, Sweden
   \newline
   (18) Dept. of Physics, Stockholm University, SE-10691 Stockholm, Sweden
   \newline
   (19) Vrije Universiteit Brussel, Dienst ELEM, B-1050 Brussels, Belgium
   \newline

}}

\baselineskip=11.6pt

\begin{abstract}
AMANDA (\textbf{A}ntarctic \textbf{M}uon \textbf{A}nd \textbf{N}eutrino \textbf{D}etector
\textbf{A}rray) is a neutrino telescope built under the southern polar icecap, and
its scope is exploring the possibility to detect high energy cosmic neutrinos
generated by powerful celestial objects where acceleration mechanisms can bring protons
up to $10^{20}$ eV. We describe the achievements and results from the AMANDA-B10
prototype and the preliminary results from the current AMANDA-II detector showing a
dramatic increase in sensitivity. An outlook on IceCube will be given.
\end{abstract}
\newpage
\baselineskip=14pt
\section{Introduction}
\label{sec:intro}

Observational astronomy and astrophysics are based on the detection of information carriers
to determine the properties of the emitting objects. Three types of carriers
have been used so far: electromagnetic radiation, cosmic rays, and neutrinos. The electromagnetic
radiation was the first to be observed. Due to their neutral charge, photons are not
deflected by interstellar and intergalactic magnetic fields, therefore they point directly
back to their source. Photons are emitted within the outer regions of the sources and they cannot
give any information about the source internal processes. Moreover, high energy photons are
affected by interactions with the microwave and infrared backgrounds (mostly through
$\gamma \gamma \rightarrow e^{+}e^{-}$), as they traverse intergalactic distances. This limits
the range of cosmological sources that can be observed with the highest energy photons.

\begin{figure}[htb]
\centerline{\psfig{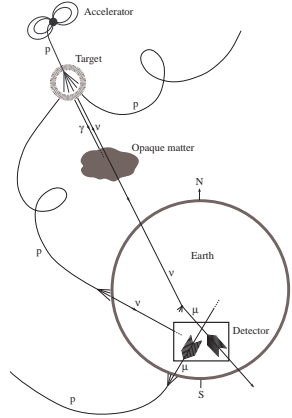}}
\caption{\it High energy information carriers}
\label{fig:mess}
\end{figure}

Cosmic rays are composed of charged nuclei and are deflected by magnetic fields, so they don't bring
directional information and they are subject to energy loss processes during their propagation, which
deforms the energy spectrum measured at the Earth with respect to the one at the source. Moreover the
highest energy cosmic rays ($E\geq 5\cdot 10^{19}$eV) are expected to be absorbed due to collisions
with microwave background photons, through $p+\gamma(2.7^{\circ}K) \rightarrow p+\pi ^{\circ}$ (Greisen-Zatsepin-Kuzmin,
or GZK cutoff \cite{gzk}). This would generate a degradation of proton energy after a distance of 50 Mpc
\cite{hires}. Thus, while cosmic rays carry some information about the energy distribution mechanisms at their source,
they cannot indicate directly where those sources are located and they cannot propagate for cosmological
distances (see figure 1).

Neutrinos have no electric charge and have a low interaction cross-section with matter. They can
propagate from the production sources through cosmological distances, un-deflected by magnetic
fields. The distance at which the Universe can be observed with high energy neutrinos is limited only by
the signal strength of the source.
Thus, they can reach the Earth carrying the complete flux and spectral information about their generation
mechanism. On the other hand, their low cross-section makes them difficult to detect. Very large detectors,
long exposure times and sophisticated data analyses are needed to measure high energy extraterrestrial
neutrino fluxes and spectra.
High energy cosmic neutrinos are beleived to be produced in energetic accelerating environments through
p-p or p-$\gamma$ interactions via $\pi$ production and decay. Such an accelerator might be the core
of an active galaxy, powered by a supermassive black hole. The energy dependence of the expected neutrino spectrum
is predicted to be $\sim$ E$^{-2}$.

After Greisen's 1960 review on cosmic ray showers \cite{greisen}, the detection of neutrinos was suggested by
Markov and Zheleznikh \cite{mar} via the process

\begin{equation}
\nu_l(\bar{\nu_l})+N \rightarrow l^{\pm} + X
\label{eq:nu}
\end{equation}

\noindent of upward or horizontal neutrinos interacting with a nucleon $N$ of the matter surrounding the detector. At high
energy, approximately half of the neutrino energy is carried by the produced lepton and the angle between the
neutrino and the lepton is, for muon flavor \cite{gaisser}

\begin{equation}
<\psi _{\nu_{\mu},\mu}> \approx \left({0.7^{\circ}\over E_{\nu}/\mathrm{TeV}}\right)^{0.7}
\label{eq:nuangle}
\end{equation}

\noindent where $E_{\nu}$ the neutrino energy.
At sufficiently high neutrino energy the muon is collinear with the parent neutrino and the muon detection
will give information on the neutrino direction. The possibility to measure the high energy cosmic
neutrino direction opened the door to neutrino astrophysics. This idea started to become reality
with the DUMAND project \cite{dumand} in the deep Pacific Ocean near Hawaii. This pioneering effort
opened the road to the realization of the first dedicated working projects, Baikal \cite{baikal}
and AMANDA \cite{amanda}, and to the two planned projects in the Mediterranean Sea NESTOR \cite{nestor} and ANTARES
\cite{antares}.

\section{AMANDA-II Description and Operational Principles}
\label{s:amanda}

\begin{minipage}[h]{6cm}
\psfig{figure=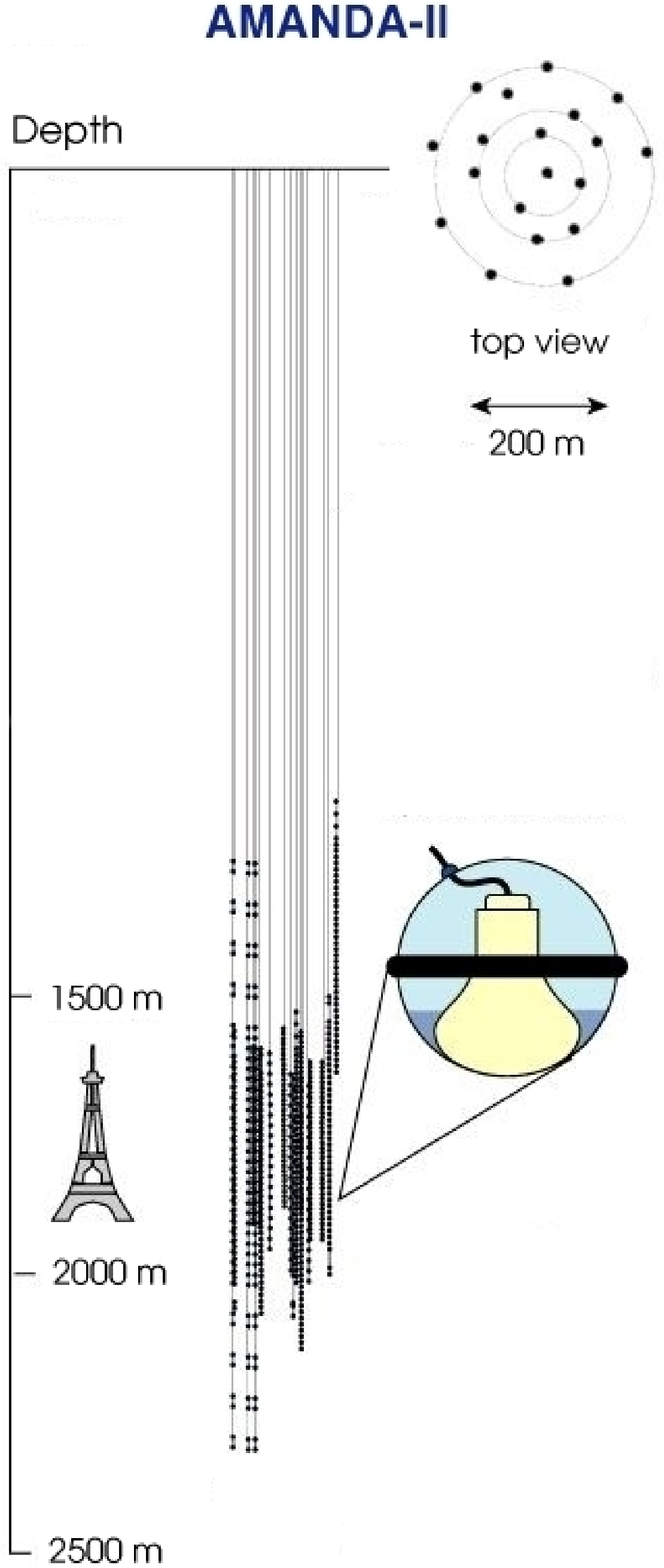,height=12cm}
Figure 2: {\it AMANDA-II schematics}
\end{minipage}
\begin{minipage}[h]{6cm}
\baselineskip=14pt
\noindent AMANDA is a neutrino telescope built and operated at the
Geographic South Pole. The final detector configuration, called AMANDA-II, was deployed in four South Pole
campaigns between November 1995 and February 2000 \cite{icrc_wis}. The detector consists of 677 optical modules (OM)
arranged on 19 vertical strings deployed at depths between 1300-2400 meters. The used instrumented volume ranges
between 1500-2000 meters, and covers a cylinder of 200 meters diameter (See figure 2).

An AMANDA-II OM consists of a single 8 inch Hamamatsu R5912-2 photomultiplier tube (PMT) housed in a glass pressure
vessel. The PMT is optically coupled to the glass housing by a transparent gel. Each module is connected to
electronics on the surface by a dedicated electrical cable, which supplies high voltage and carries the anode signal
of the PMT. For each event, the OM is read out by a peak-sensing ADC and TDC capable of registering up to eight separate
pulses \cite{atmnub10}. The first 10 strings, completed in 1997 and containing 302 OMs, are known as AMANDA-B10, with
120 m diameter.
\end{minipage}
\vskip .2cm

\begin{figure}[htb]
\psfig{figure=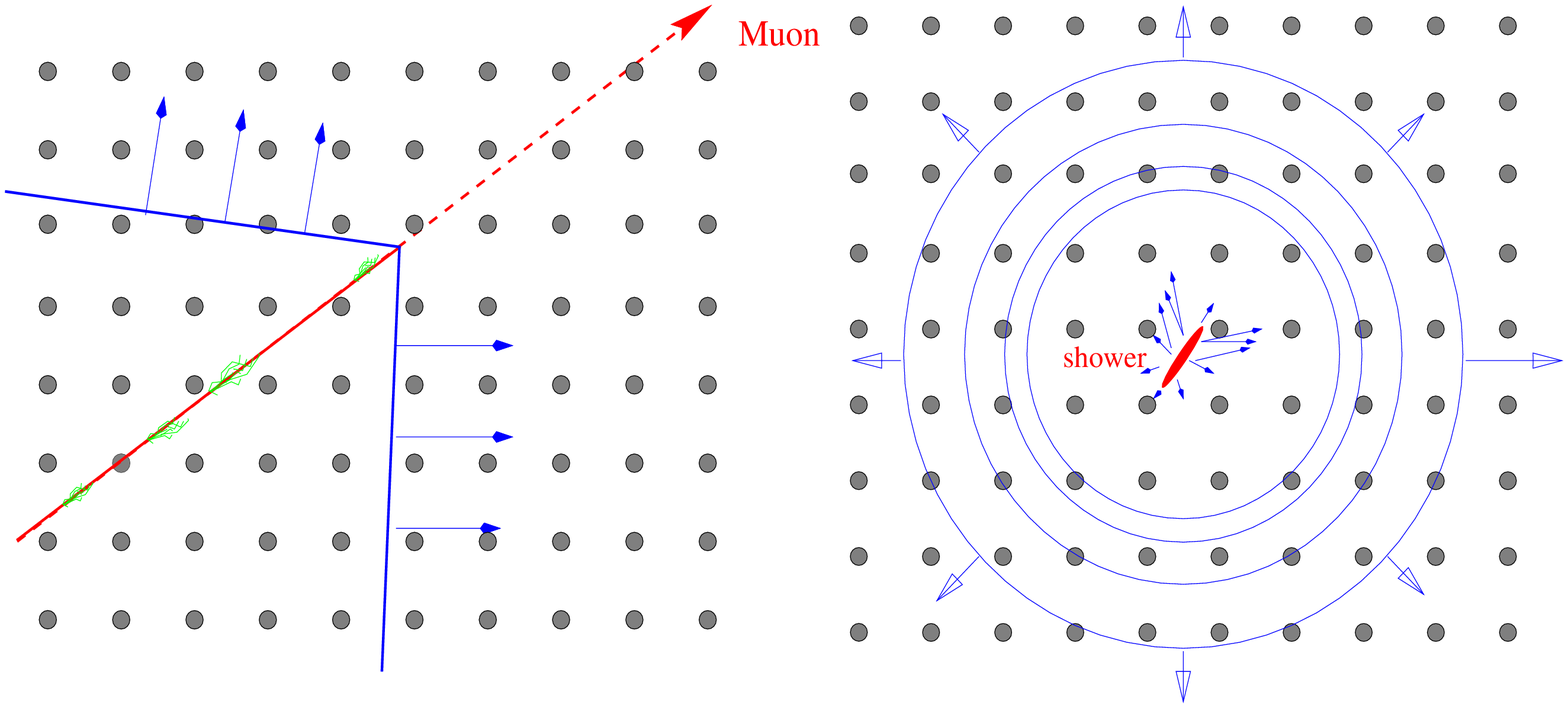,width=12.5cm}
Figure 3: {\it Cherenkov light detection topology in the AMANDA-II array for tracks (left) and cascades (right)}
\end{figure}

The holes where AMANDA strings are buried, were produced with high pressure hot water ($\sim 90^{\circ}C$). The
drilling procedure was continuously monitored and the average hole diameter was approximately $50$ cm over the entire hole,
with a depth-variation, due to the ice temperature profile.
The string deployment is done just after drilling is finished and before the hole diameter starts to shrink due to
refreezing. Each OM is connected to the main cable and tested in situ at deployment time. The South Pole Air Shower
Experiment (SPASE) is a surface array which measures the the shower direction and size, via the electron density
\cite{spase}. The SPASE/AMANDA coincident events are used to study the cosmic ray composition above $10^{15}$ eV.

The channels through which neutrino telescopes detect neutrinos above energies of a few tens of GeV is by observing
the Cherenkov light from relativistic leptons produced in $\nu_l$-nucleon interactions (eq. \ref{eq:nu}) in or near the telescope.
The primary channel is the $\nu_{\mu}$, which leads to a high energy $\mu$ track. The muons can propagate several
kilometers through the ice and they can be reconstructed with reasonable precision even with a coarsely instrumented
detector, provided the medium is sufficiently transparent.
The time and amplitude of detected signals are used to reconstruct the path of the muon through the detector.

The other possible detection channels $\nu_e$ and $\nu_{\tau}$ have a different behavior. An electron, produced by
a $\nu_e$, will generate an electromagnetic cascade, which is confined to a volume of a few cubic meters. It overlays
the hadronic cascade of the primary interaction vertex. The size of both cascades is small compared to the OM spacing
in the array and the signature is an expanding spherical shell of Cherenkov photons (figure 3). A $\tau$, produced by a
$\nu_{\tau}$, will decay immediately and generate a second hadronic cascade. However, at energies above 1 PeV the two
cascades are separated by several tens of meters, connected by a single track.

A detected event corresponds to a realization of a majority trigger of a predefined hit OM channels within 2.5 $\mu$s.
To ensure that the observed muons are produced by neutrinos, the Earth is used as a filter and only up-going
muons are selected. This condition is used to reduce the contamination from down-going cosmic ray muon flux.
The location depth in the ice serves to further minimize the flux of cosmic ray muons.

\begin{figure}
\centering
\begin{minipage}[c]{0.46\textwidth}
\psfig{figure=scatter.epsi,height=5.2cm}
\end{minipage}
\hspace{0.8cm}
\begin{minipage}[c]{0.46\textwidth}
\psfig{figure=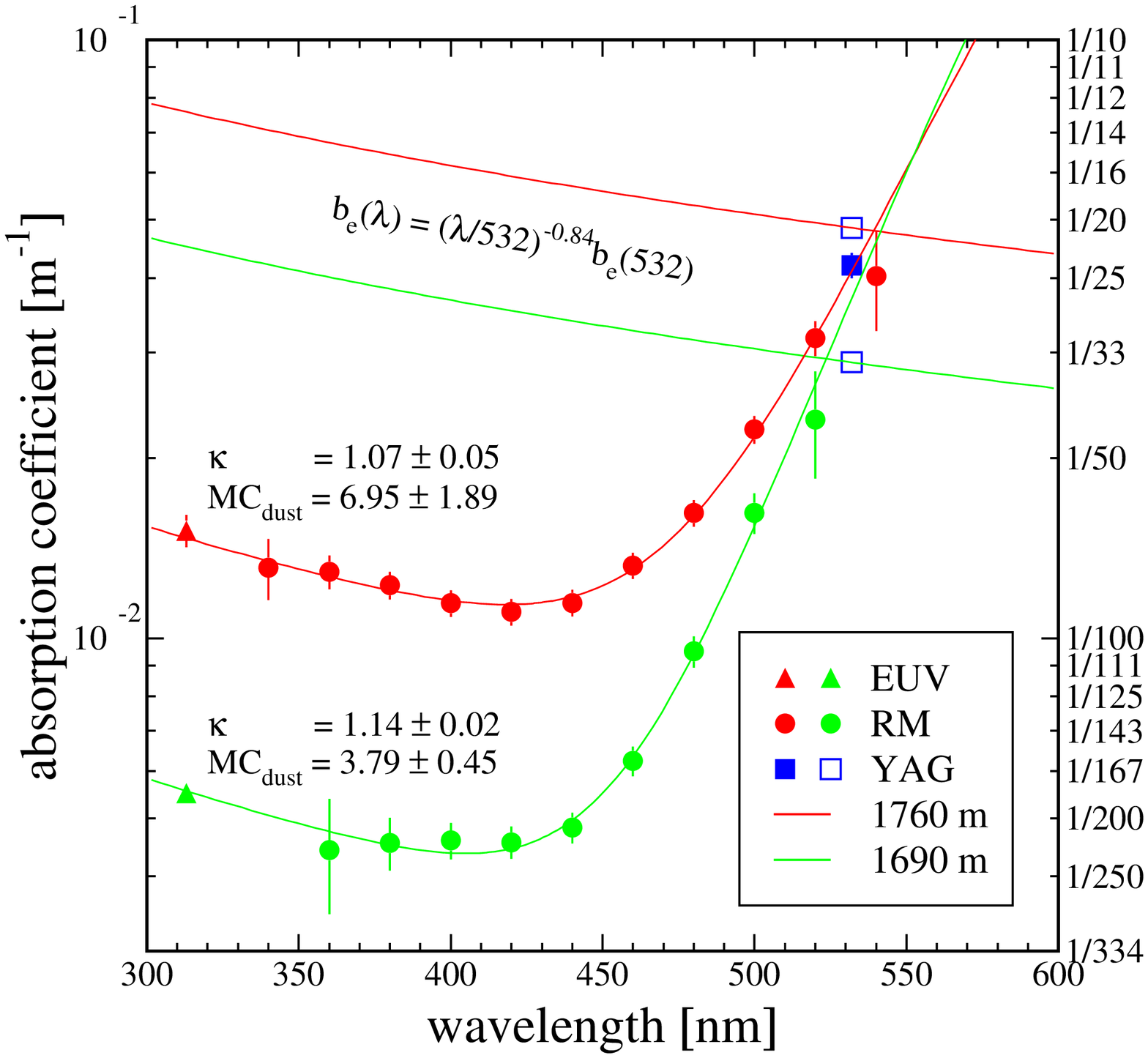,height=5.2cm}
\end{minipage}
\vskip .4cm
\begin{minipage}[t]{\textwidth}
Figure 4: {\it Left: effective scattering coefficient as a function of depth at 532 nm wavelength.
               Right: absorption and effective scattering ($b_e$) coefficients versus wavelengths.}
\end{minipage}
\end{figure}
%\vskip .4cm

The ice surrounding AMANDA is part of the neutrino telescope and its optical properties are
very important and are directly connected to the detector sensitivity. These properties were
studied in detail, using in-situ light emitters located on the strings and the down-going muon flux itself. This
study shows the ice is not homogeneous \cite{oprop}, but it can be considered as made of several horizontal
layers, laid down by varying climatological conditions in the past \cite{buford}. Figure 4 shows the effective
scattering length versus depth for 532 nm. The figure shows also the absorption and effective scattering lengths as
a function of wavelength for 2 different depths. At 400 nm (corresponding to the maximum OM optical sensitivity) the
average absorption length at AMANDA-II depth is 110 m and the average effective scattering length is 20 m \cite{atmnub10}.

The light emitters on the strings are used also as a detector time and geometry calibration system.
The time calibration is a measure of the cable length and, therefore, of PMT signal time delay due to cable
propagation. The overall precision on the photon arrival times is approximately 5 ns. The position calibration
is performed using two sets of information: the absolute array position, obtained from drill recordings
and pressure sources, and the relative positions, obtained with the measured propagation times of photons between 
different light emitters and receivers. The position of the OMs are determined with a precision of about 0.5
m \cite{b4}.

\section{Simulation and Data Processing}
\label{s:datasim}

In order to study the detector response to the different event sources, a Monte Carlo simulation covering the complete
chain from the primary particle flux interactions to the detector response, is used to evaluate the array sensitivity.
The down-going muon flux is generated by using CORSIKA (v6.020) shower generator with QGSJET01 interaction model
\cite{corsika}; the up-going muons are generated with a program called NUSIM \cite{hill}, which allows us to simulate
any $\nu_{\mu,e}$ spectrum, the propagation through the Earth and their interactions. The produced muons
are propagated through the ice, taking into account all the relevant energy losses \cite{mmc}. The Cherenkov photon
production and propagation are simulated taking into account the ice optical properties and the OM acceptance.
The detector trigger simulation includes the hardware response of each OM \cite{amasim}. At this stage the simulated
data are treated exactly as the experimental raw data and passed through the same data processing.

The AMANDA data are passed through a standard filter which is common to all the neutrino analyses. This
filtering is performed in 2 separate levels. The level 1 filter (L1) includes the OM cleaning, which excludes the channels
which are dead or have odd transient behavior, performs time calibration and a first guess reconstruction. A loose angular
cut on the approximate reconstructed muon track is used to filter upward moving tracks and to reduce the amount of data
to be passed to the following level. The level 2 filter (L2) applies iterative likelihood reconstructions, which take into
account the ice properties, and a Bayesian reconstruction \cite{reco}. A second cut on the likelihood reconstructed
angle is applied to further reduce the down-going event contamination.
To study down-going muon events the same filter levels are applied but with no angular cut.

\section{AMANDA-II Background Sources}
\label{s:back}

\subsection{Atmospheric Muons}
\label{ss:atmu}

The largest background source for AMANDA is represented by the down-going cosmic ray muon flux, which is
5 orders of magnitude larger than the total expected neutrino flux, and contributes to the detector trigger
with a rate of $\sim 100$ Hz. Selecting upward reconstructed events
(sect. \ref{s:datasim}) has an important effect in removing a large fraction of these events, but due to the finite
reconstruction resolution a small fraction of down-going muon events is mis-reconstructed as upward.
Such events represent a background which still needs to be eliminated in any neutrino analysis. Nevertheless
the large cosmic ray muon flux can be used to study the detector response, possible systematic effects and also
to perform physics measurements. The cosmic ray muon angular and vertical intensities were measured in
AMANDA-II and they agree well with simulation, with other experimental measurements and with theoretical
calculations \cite{icrcmu}. These measured distributions are related to the sea-level muon and primary Cosmic
Ray energy spectrum. The measured spectral index of sea-level differential muon spectrum ($\gamma_{\mu}=2.70\pm 0.04$)
and of primary differential cosmic ray spectrum ($\gamma_{prim}=3.72\pm 0.17$) are in agreement with other experimental results
\cite{heiko}. The down-going muon events which are in coincidence with SPASE surface array \cite{spase} were used to measure the
cosmic ray mass composition as a function of the primary energy near the knee. An energy resolution of $\sim 7\%$
was achieved in the energy range of 1-10 PeV and a tendency toward heavier nuclei above the knee was measured \cite{comp}.

The down-going muon flux is well understood and an on-going analysis is examining the sensitivity for detecting
muons from charmed meson decays in the atmosphere. 

Uncorrelated, multiple atmospheric muon events represent another source of background for neutrino analyses. The expected
rate of such events is very low in AMANDA-II (of the order of $\sim 1$ Hz, higher than
neutrino trigger rate) but their contamination cannot be neglected as for AMANDA-B10. The contamination of such
events becomes clear at high quality cut level where the bulk of atmospheric muons has been eliminated. A typical
uncorrelated muon event could be a muon hitting the upper detector and one hitting the lower detector shortly before.
The reconstruction tries to find the best up-going single muon track compatible with the recorded hit times. The easiest way to
eliminate such a background is to require a uniform hit distribution along the reconstructed track.

\begin{figure}
\centering
\begin{minipage}[c]{0.46\textwidth}
\psfig{figure=fig_18_zenith_q7_25bins_adb2.epsi,height=5cm}
\end{minipage}
\hspace{0.1cm}
\begin{minipage}[c]{0.46\textwidth}
\psfig{figure=fig_25_skyplot_overlap_radec.epsi,height=3.1cm}
\end{minipage}
\vskip .4cm
\begin{minipage}[t]{\textwidth}
Figure 5: {\it Left: the zenith angle distribution of up-going reconstructed events.
               Monte Carlo prediction is normalized to the experimental data.
               Right: skyplot of the reconstructed events in equatorial coordinates.}
\end{minipage}
\end{figure}

\subsection{Atmospheric Neutrinos}
\label{ss:atnu}

In order to measure the atmospheric neutrino component, a rejection factor of $\sim 10^5$ is needed to eliminate the
cosmic ray muon background contamination. The detection of atmospheric neutrinos in AMANDA provided the first check
of the detector sensitivity \cite{b4,atmnub10}.

Figure 5 shows the published result from the year 1997 of AMANDA-B10. The 130.1 days livetime analyzed gave
$N_{exp}=204$ experimental reconstructed events with a prediction of $N_{MC}=279\pm 3$. Normalizing the
Monte Carlo simulated events to the experimental sample at high quality level gives a background contamination
(i.e. a residual leakage of down-going muons reconstructed as up-going) of 5-10 \%. The low neutrino
acceptance close to the horizon reflects the effect of down-going muon background elimination. The track pointing
resolution is $\psi=3.2^{\circ}$ and the skyplot, shown in figure 6, is consistent with a uniform isotropic flux.
The estimated energy of detected neutrino sample ranges between 66 GeV and 3.4 TeV (90 \% of events) \cite{atmnub10}.

AMANDA-II detector, with its increased size, has a better sensitivity than AMANDA-B10 and its larger diameter allows us to
better reconstruct the horizontal events. Thus the acceptance near the horizon is improved and the down-going
muon flux rejection turns out to be an easier task than with the smaller detector. Requiring a minimum number of
direct hits (i.e. hits with a small time residual with respect to the direct Cherenkov light time of propagation),
a high likelihood of reconstruction value and a uniform light deposition along the reconstructed track, we record
about 4 well reconstructed neutrino events per day in AMANDA-II. The preliminary unfolded $\nu_{\mu}$ sea-level energy
spectrum was determined and is in agreement with the expectation up to $10^4$ GeV \cite{nunfold}.

\section{AMANDA Analyses}
\label{s:analyses}

\subsection{Diffuse Flux of $\nu_{\mu}$}
\label{ss:diffuse}

A search for high energy neutrino flux from unresolved sources throughout the Universe was done in AMANDA-B10,
using the data taken during 1997.
After cleaning the experimental data sample from atmospheric muon background, the neutrino event selection
was designed to retain high energy track-like events \cite{diffb10} in order to eliminate the lower energy
atmospheric neutrino background. The energy cut is performed through the hit channel multiplicity observable.

The AMANDA-B10 sensitivity \footnote{defined as 90\% CL average upper limit from an ensemble of identical
experiments with no signal present} was tested with respect to an assumed high energy neutrino spectrum given by
$\Phi_{\nu_{\mu}}=10^{-5}\cdot $E$^{-2}~$cm$^{-2}$s$^{-1}$sr$^{-1}$GeV$^{-1}$.
This spectrum was simulated along with the atmospheric neutrinos and the detector response was studied.
Optimizing the atmospheric neutrino background rejection with a suitable channel multiplicity cut, there is no
excess of events with respect to the estimated background and a limit was set on the high energy flux.
Figure 6 shows the effect of the energy cut on the atmospheric and high energy neutrino events (on the left)
and the calculated limit (on the right). The limit, including 25\% atmospheric neutrino flux systematic uncertainty, is
$\Phi^{lim}_{\nu_{\mu}}<8.4\times 10^{-7}\cdot $E$^{-2}~$cm$^{-2}$s$^{-1}$sr$^{-1}$GeV$^{-1}$

As stated in section \ref{s:back}, AMANDA-II is bigger than AMANDA-B10 and the background which needs to be eliminated
has larger intensity and includes the coincident uncorrelated atmospheric muon events. The tested high energy neutrino
flux is $\Phi_{\nu_{\mu}}=10^{-6}\cdot $E$^{-2}~$cm$^{-2}$s$^{-1}$sr$^{-1}$GeV$^{-1}$ and the optimized energy cut selects events
between 10 TeV and 10 PeV. Since 1997 about 10 times the exposure has been achieved in total with AMANDA-B10 (1997-99)
and AMANDA-II (2000-present). This combined data set has a limit-setting potential more than 5 times better than the above
AMANDA-B10 limit.

\begin{figure}
\centering
\begin{minipage}[c]{0.46\textwidth}
\psfig{figure=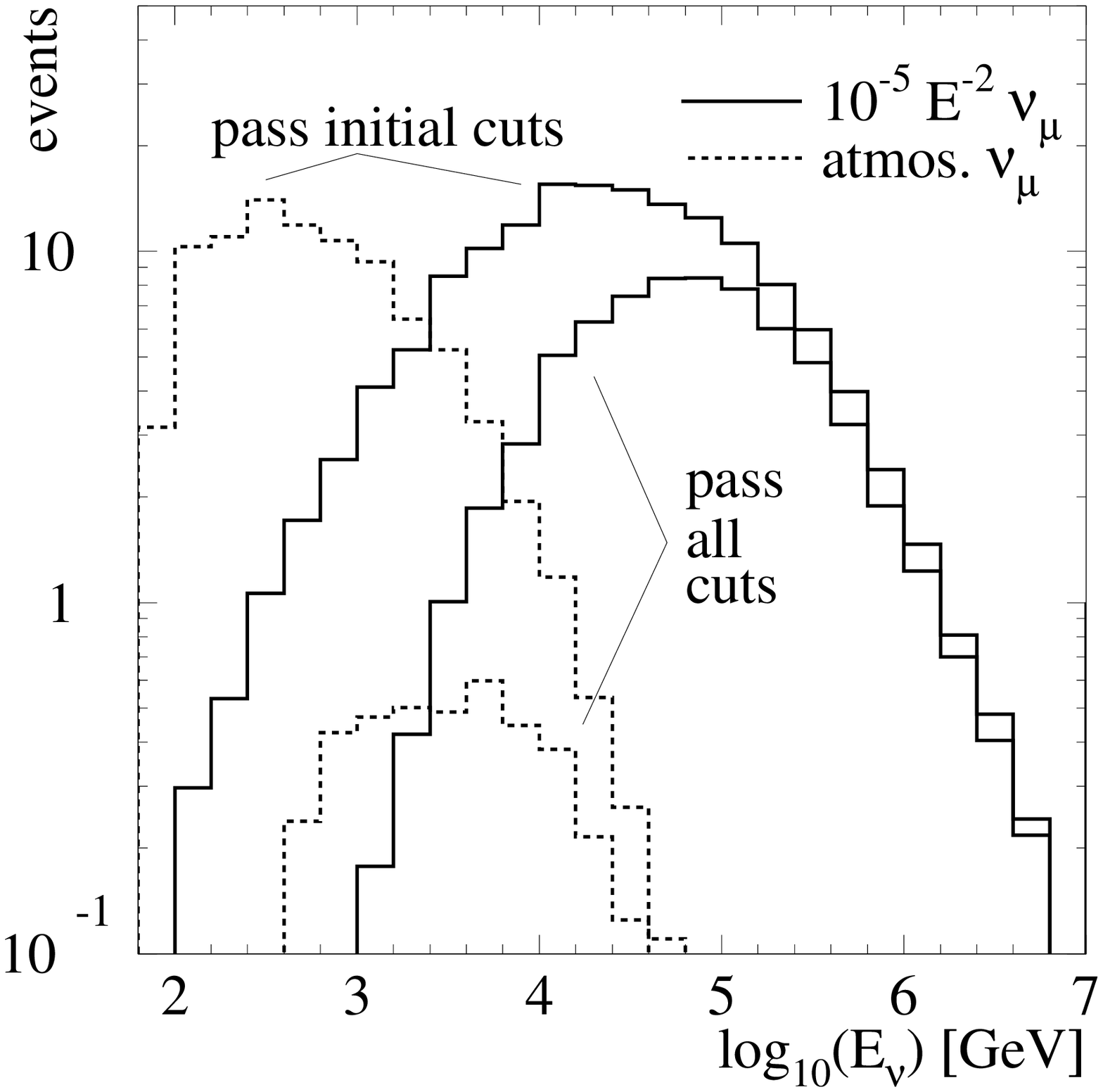,height=5.5cm}
\end{minipage}
\hspace{0.5cm}
\begin{minipage}[c]{0.46\textwidth}
\psfig{figure=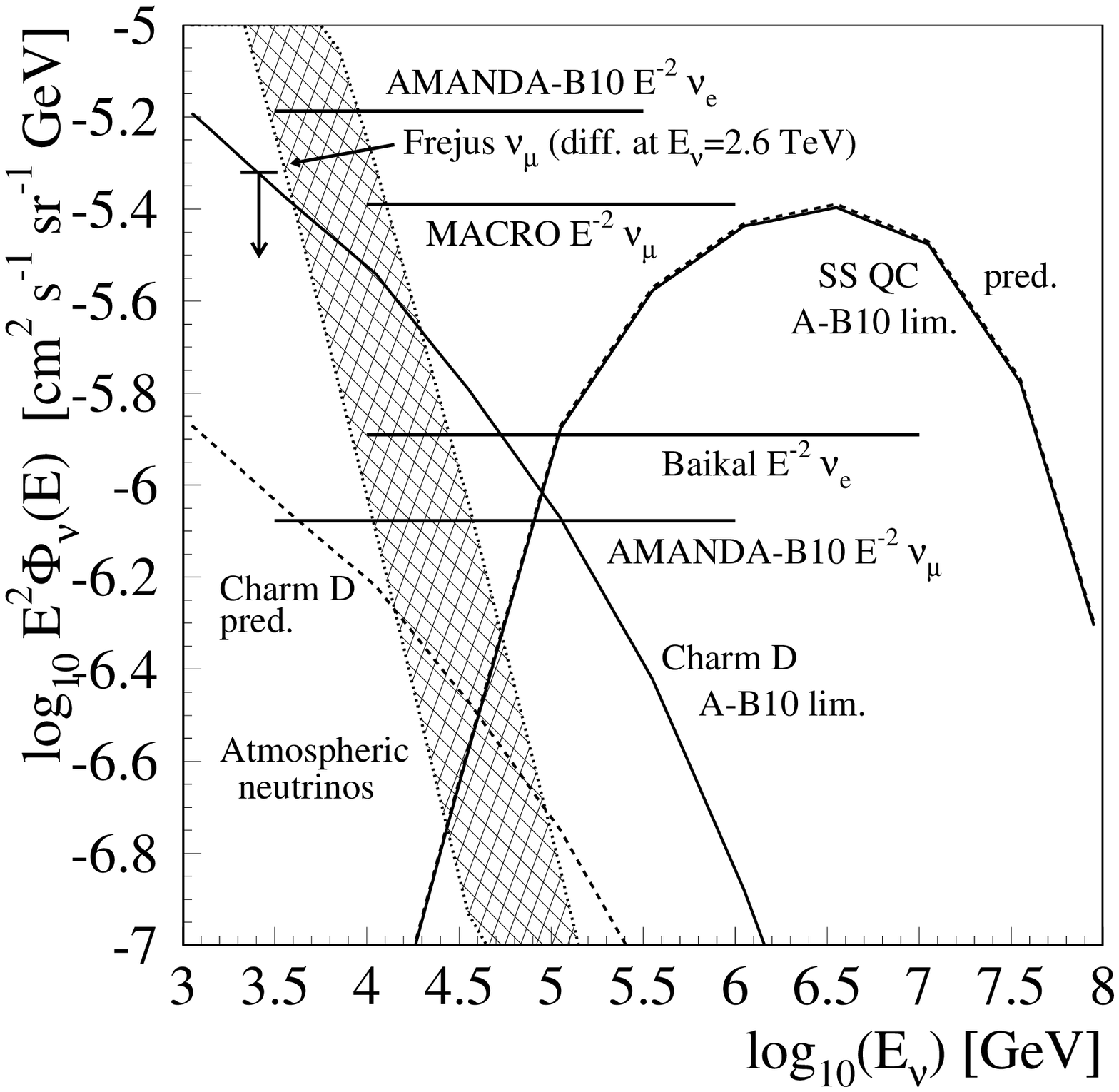,height=5.5cm}
\end{minipage}
\vskip .4cm
\begin{minipage}[t]{\textwidth}
Figure 6: {\it Left: energy spectrum of the incident atmospheric (dashed line) and E$^{-2}$ (solid line) neutrinos
               for events that pass the initial cuts, and after channel multiplicity cut.
               Right: Summary of experimental 90\% classical confidence level flux limits from various detectors
               assuming an E$^{-2}$ spectrum.}
\end{minipage}
\end{figure}

\begin{figure}
\centering
\begin{minipage}[c]{0.46\textwidth}
\psfig{figure=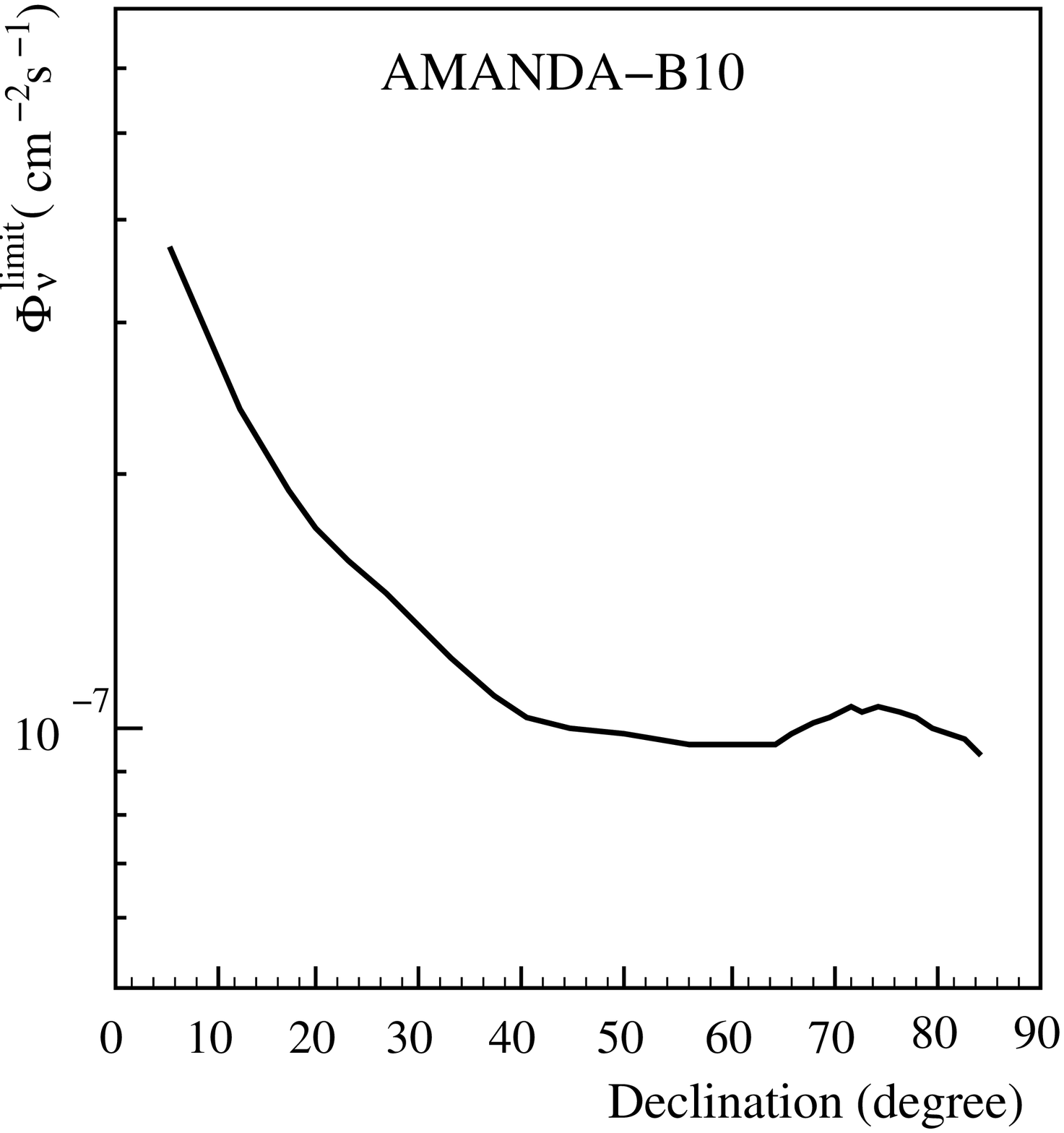,height=4.2cm,width=5.5cm}
\end{minipage}
\hspace{0.2cm}
\begin{minipage}[c]{0.46\textwidth}
\psfig{figure=sens-v-dec-syst2.epsi,height=4.2cm}
\end{minipage}
\vskip .4cm
\begin{minipage}[t]{\textwidth}
Figure 7: {\it Left: AMANDA-B10 upper 90\% CL limit on neutrino flux as a function of declination.
               Right: AMANDA-II sensitivity as a function of declination.}
\end{minipage}
\end{figure}

\subsection{Point Sources of $\nu_{\mu}$}
\label{ss:pointnu}

If the dominant extraterrestrial neutrino flux is emitted by a few particularly bright or close sources, their direction
could be resolved. For that reason the 1997 data taken with AMANDA-B10 were analyzed also to search for neutrinos
from point sources \cite{pointb10}. The sky was divided in 154 search bins, with half width of about $5^{\circ}$. Due
to the lower background within each search bin, it is possible to relax the cuts, with respect to the search for
diffuse flux of neutrinos, and thus gain a higher signal efficiency. The pointing resolution results to be
$\psi=3.9^{\circ}$. The probability for the observed number of events in each search bin shows no significant excess
with respect to the pure background expectation. Therefore a limit was derived. Figure 7 shows, on the left, the upper
limit (90\% CL) as a function of declination.

The search for high energy neutrinos from point sources was also performed in AMANDA-II using the data taken during 2000 \cite{pointaii}.
Due to the better pointing resolution ($\psi=2.3^{\circ}$) of AMANDA-II, in this analysis the sky was divided in 300 search bins,
with half width of about $3.5^{\circ}$. No significant excess of events was observed in any search bin. Figure 7, on the right, shows the
AMANDA-II sensitivity (integrated over neutrino energies above 10 GeV) as a function of declination. It is evident that AMANDA-II
has a more uniform angular sensitivity and, in average, about a factor of 5 better than AMANDA-B10.

\begin{table}[t]
  \centering
  \caption{\it Preliminary 90\% CL upper limits on a selection of candidate sources in AMANDA-II. The limits
               $\Phi_{\mu,\nu}^{lim}$ are calculated for an assumed E$^{-2}$ spectral shape, integrated
               above E$_{\mu,\nu}$ = 10 GeV and in units of $10^{-15}cm^{-2}s^{-1}$ and $10^{-8}cm^{-2}s^{-1}$,
               respectively.}
  \vskip 0.1 in
  \begin{tabular}{|c|r|r|r|r|r|r|}
    \hline
    Candidate    & Dec ($^{\circ}$) & R.A. (h) & $n_{obs}$ & $n_b$ & $\Phi_{\mu}^{lim}$ & $\Phi_{\nu}^{lim}$ \\
    \hline
    \hline
    SS433        &  5.0             & 19.20   & 0          & 2.38  & 0.8                & 0.6                \\
    M 87         & 12.4             & 12.51   & 0          & 0.95  & 1.1                & 0.9                \\
    Crab Nebula  & 22.0             &  5.58   & 2          & 1.76  & 2.1                & 2.1                \\
    Mkr 421      & 38.2             & 11.07   & 3          & 1.50  & 2.6                & 3.1                \\
    Mkr 501      & 39.8             & 16.90   & 1          & 1.57  & 1.3                & 1.6                \\
    Cygnus X-3   & 41.0             & 20.54   & 3          & 1.69  & 2.5                & 3.1                \\
    QSO 0219+428 & 42.9             &  2.38   & 1          & 1.63  & 1.1                & 1.4                \\
    Cassiopea A  & 58.8             & 23.39   & 0          & 1.01  & 0.7                & 1.1                \\
    QSO 0716+714 & 71.3             &  7.36   & 2          & 0.74  & 2.4                & 3.8                \\
    \hline
  \end{tabular}
  \label{extab}
\end{table}

Table \ref{extab} shows the preliminary 90\% CL upper limits on a selection of candidate sources \cite{pointaii}.
$n_{obs}$ is the number of observed events within the search bin and $n_b$ the number of expected background,
determined by measuring the events off-source in the same declination band. These upper limits are substantially better
than the limits from AMANDA-B10 \cite{pointb10}.

\subsection{Flux of $\nu_{\mu}$ from Gamma Ray Bursts}
\label{ss:grbnu}

AMANDA detector data were analyzed to search high energy neutrinos spatially and temporally coincident with 317
triggered GRBs, detected by the BATSE satellite detector, and 153 non-triggered GRBs, obtained by searching the BATSE archived
data \cite{grbnu}. The experimental data are from AMANDA-B10 (taken in 1997-99) and AMANDA-II (taken in 2000)
combined. The preliminary results are consistent with no GRB neutrino signal. A 4-year combined neutrino event
90\% CL upper limit of 1.45 for the 317 BATSE-triggered bursts was derived.

\subsection{Flux of Ultra High Energy $\nu_{\mu}$}
\label{ss:uhenu}

AMANDA-B10 data were searched for muon neutrinos with energies above $10^{16}$ eV. At these energies the Earth is
opaque to neutrinos and the events are concentrated at the horizon. The background is represented by the large
muon bundles from down-going atmospheric events. The absence of extremely bright events in excess with respect to
the expected background, allowed us to put a limit of
$\Phi^{lim}_{\nu_{\mu}}<7.2\times 10^{-7}\cdot $E$^{-2}$cm$^{-2}$s$^{-1}$sr$^{-1}$GeV$^{-1}$ in the energy range
$2.5\times 10^{15}eV<E_{\nu}<5.6\times 10^{18}eV$ \cite{uheb10}. The preliminary sensitivity of AMANDA-II, for one
year of operation, is about a factor of three better.

\subsection{Diffuse Flux of $\nu$ of all Flavors}
\label{ss:casc}

As explained in section \ref{s:amanda}, the signature for $\nu_{e,\tau}$ is different from the one for $\nu_{\mu}$,
therefore different fit procedures need to be used to reconstruct such events. The advantage of detecting these events
is the lower atmospheric background, the better energy resolution with respect to the muon case, since the full energy
is deposited in the detector, and the sensitivity to all neutrino flavors, because of neutral current interactions.
Detection of all the three flavors increases the sensitivity, since, for typical astrophysical fluxes, the ratio of
neutrino flavors $\phi_{\nu_e}:\phi_{\nu_{\mu}}:\phi_{\nu_{\tau}}\approx 1:2:0$, becomes $1:1:1$ at the detection site, due to neutrino
oscillations. For contained cascades, the cascade vertex position resolution is $\Delta r \sim$ 5 m for both AMANDA-B10
and AMANDA-II \cite{cascb10} and the energy resolution is $\Delta (Log E)\sim$ 0.1-0.2. The bigger AMANDA-II volume
increases the sensitivity to higher cascade energies and the angular acceptance is nearly uniform over $4\pi$ \cite{cascaii}.

Figure 8 shows the effective volume of AMANDA-II as a function of neutrino energy for all neutrino flavors, and the upper
limits for AMANDA-II compared with those of AMANDA-B10. An order of magnitude improvement in sensitivity is evident
from the figure.

\begin{figure}
\centering
\begin{minipage}[c]{0.46\textwidth}
\psfig{figure=veff.epsi,height=5.2cm}
\end{minipage}
\hspace{0.2cm}
\begin{minipage}[c]{0.46\textwidth}
\psfig{figure=limit.epsi,height=5.2cm}
\end{minipage}
\vskip .4cm
\begin{minipage}[t]{\textwidth}
Figure 8: {\it Left: Effective Volume versus neutrino energy for AMANDA-II.
               Right: Upper 90\% CL limits on the flux of cosmic neutrinos following a $E^{-2}$ spectrum}
\end{minipage}
\end{figure}

\subsection{Flux of $\nu_e$ from Supernov{\ae}}
\label{ss:snnu}

AMANDA-B10 data were also searched for the detection of low energy ($\sim$ MeV) neutrinos from Supernov{\ae} \cite{snnu}
through the process $\bar{\nu}_e+p\rightarrow n+e^{+}$. The $e^{+}$ track length is too short to be reconstructed, but the
Cherenkov light from many such tracks can induce a correlated increase in the noise rate of the OMs that can be
detected. The OM count rate was monitored continuously in 10 s bins in search for such an excess.
The analysis includes a specific OM selection criterion, which keeps only modules with very
stable behavior over the years 1997-98. A stable counting rate is achieved with a moving average. AMANDA-B10
yields a 70\% coverage of the Galaxy with one backgound fake per year with 90\% efficiency. AMANDA-II supernova
data are being analyzed now and indicate a 90\% coverage of the Galaxy.

\subsection{Flux of $\nu_{\mu}$ from WIMPs}
\label{ss:wimps}

An indirect search for non-baryonic dark matter, in the form of Weakly Interacting Massive Particles, from the
center of the Earth was done in AMANDA-B10 using the data taken in 1997-99 \cite{wimps}. This analysis requires
the selection of well reconstructed vertically up-going events. An iterative discriminant analysis retains
30\% of the signal (represented by neutralino annihilations) and 0.3\% of background. The observation of no excess
with respect to the expected background was interpreted as a limit on the muon flux from neutralino annihilation.
An upper limit of $\sim 10^3~\mu~km^{-2}yr^{-1}$ for $m_{\chi}>10^3$ GeV, and less stringent values for smaller
neutralino masses, were derived from AMANDA-B10. Preliminary analyses show that AMANDA-II has competitive sensitivity
to search also for WIMPS from the Sun.

\begin{figure}
\centering
\begin{minipage}[c]{0.46\textwidth}
\psfig{figure=significances.e2.diff.epsi,height=5cm}
\end{minipage}
\hspace{0.2cm}
\begin{minipage}[c]{0.46\textwidth}
\psfig{figure=significances.e2.point.lin.epsi,height=5cm}
\end{minipage}
\vskip .4cm
\begin{minipage}[t]{\textwidth}
Figure 9: {\it IceCube sensitivity to diffuse neutrino fluxes (on the left) and to point sources of
                neutrinos (on the right), as a function of exposure. P(5$\sigma$)=x\% means that the probability that
                a signal shows up as a 5 sigma effect is x\%.}
\end{minipage}
\end{figure}

\section{The Future: IceCube}
\label{s:icecube}

IceCube is the kilometer-cubed successor of AMANDA-II. To be built starting in 2004, it will consist of 4800
10-inch PMTs enclosed in transparent pressure spheres, distributed along 80 vertical strings up to a depth
of 2400 meters \cite{icecube}. As a significant improvement over the AMANDA technology, each OM will also
house the electronics to digitize the PMT pulses, retaining the full waveform information. The digitized
information is sent to the surface processors, which form a global trigger. Triggered events will be filtered
and reconstructed on-line, and the relevant information will be transmitted via satellite to the northern
hemisphere.

Figure 9 shows the expected IceCube sensitivities for diffuse and point source $\nu_{\mu}$ fluxes as a function
of detector exposure.

\section*{Acknowledgments}

This research was supported by the following agencies: National Science Foundation Office of Polar Programs,
National Science Foundation--Physics Division, University of Wisconsin Alumni Research Foundation, USA;
Swedish Research Council, Swedish Polar Research Secretariat, and Knut and Alice Wallenberg Foundation,
Sweden; German Ministry for Education and Research, Deutsche Forschungsgemeinschaft (DFG), Germany; Fund
for Scientific Research (FNRS-FWO), Flanders Institute to encourage scientific and technological research in
industry (IWT), and Belgian Federal Office for Scientific, Technical and Cultural affairs (OSTC), Belgium.

\end{document}